\documentclass[apjl]{myemulateapj}

\usepackage{graphicx}
\usepackage{ifpdf}
\usepackage{ifthen}

\newboolean{emulateapj}
\setboolean{emulateapj}{true}

\newcommand{\ctbd}[1]{}

\newcommand{\lc}{light curve}
\newcommand{\lcs}{light curves}

\newcommand{\cfa}{Harvard-Smithsonian Center for Astrophysics (CfA)}

\newcommand{\kms}{\ensuremath{\rm km\,s^{-1}}}
\newcommand{\ms}{\ensuremath{\rm m\,s^{-1}}}
\newcommand{\mss}{\ensuremath{\rm m\,s^{-2}}}
\newcommand{\gcmc}{\ensuremath{\rm g\,cm^{-3}}}

\newcommand{\rhk}{\ensuremath{R^{\prime}_{HK}}}
\newcommand{\logrhk}{\ensuremath{\log\rhk}}

\newcommand{\hd}[1]{\mbox{HD #1}}

\newcommand{\teff}{\ensuremath{T_{\rm eff}}}
\newcommand{\logg}{\ensuremath{\log{g}}}
\newcommand{\vsini}{\ensuremath{v \sin{i}}}
\newcommand{\feh}{[Fe/H]}

\newcommand{\rsun}{\ensuremath{R_\sun}}
\newcommand{\msun}{\ensuremath{M_\sun}}
\newcommand{\lsun}{\ensuremath{L_\sun}}

\newcommand{\rstar}{\ensuremath{R_\star}}
\newcommand{\mstar}{\ensuremath{M_\star}}

\newcommand{\astar}{\ensuremath{a_\star}}
\newcommand{\loglstar}{\ensuremath{\log{L_\star}}}

\newcommand{\mearth}{\ensuremath{M_\earth}}

\newcommand{\rpl}{\ensuremath{R_{p}}}
\newcommand{\mpl}{\ensuremath{M_{p}}}

\newcommand{\rhopl}{\ensuremath{\rho_{p}}}
\newcommand{\ipl}{\ensuremath{i_{p}}}

\newcommand{\gpl}{\ensuremath{g_{p}}}

\newcommand{\rjup}{\ensuremath{R_{\rm J}}}
\newcommand{\mjup}{\ensuremath{M_{\rm J}}}

\newcommand{\rjuplong}{\ensuremath{R_{\rm Jup}}}
\newcommand{\mjuplong}{\ensuremath{M_{\rm Jup}}}

\newcommand{\mplsini}{\ensuremath{\mpl\sin i}}

\newcommand{\figr}[1]{Fig.~\ref{fig:#1}}
\newcommand{\secr}[1]{\mbox{\S\ \ref{sec:#1}}}

\newcommand{\tabr}[1]{\mbox{Table~\ref{tab:#1}}}

\newcommand{\flwof}{\mbox{FLWO 1.2 m}}

\defcitealias{kovacs05}{KBN05}
\defcitealias{fortney06}{FMB06}

\newcommand{\hdcur}{\hd{147506}}
\newcommand{\hdcurb}{\hd{147506b}}

\shorttitle{A Super-Massive Transiting Planet}
\shortauthors{Bakos et al.}

\begin{document}
\ifthenelse{\boolean{emulateapj}}{
\title{\hd{147506}\lowercase{b}: A Super-Massive Planet in an Eccentric Orbit 
	Transiting a Bright Star\altaffilmark{$\dagger$}}}
{\title{\hd{147506}\lowercase{b}: A Super-Massive Planet in an Eccentric Orbit 
	Transiting a Bright Star\altaffilmark{\dagger}}}
\author{
	G.~\'A.~Bakos\altaffilmark{1,2},
	G.~Kov\'acs\altaffilmark{3},
	G.~Torres\altaffilmark{1},
	D.~A.~Fischer\altaffilmark{4},
	D.~W.~Latham\altaffilmark{1},
	R.~W.~Noyes\altaffilmark{1},
	D.~D.~Sasselov\altaffilmark{1},
	T.~Mazeh\altaffilmark{5},
	A.~Shporer\altaffilmark{5},
	R.~P.~Butler\altaffilmark{6},
	R.~P.~Stefanik\altaffilmark{1},
	J.~M.~Fern\'andez\altaffilmark{1},
	A.~Sozzetti\altaffilmark{1,7},
	A.~P\'al\altaffilmark{8,1},
	J.~Johnson\altaffilmark{9},
	G.~W.~Marcy\altaffilmark{9},
	J.~Winn\altaffilmark{10},
	B.~Sip\H{o}cz\altaffilmark{8,1},
	J.~L\'az\'ar\altaffilmark{11},
	I.~Papp\altaffilmark{11} \&
	P.~S\'ari\altaffilmark{11}
}

\altaffiltext{1}{\cfa,
	60 Garden Street, Cambridge, MA 02138, USA; gbakos@cfa.harvard.edu.}
\altaffiltext{2}{Hubble Fellow.}
\altaffiltext{3}{Konkoly Observatory, Budapest, P.O.~Box 67, H-1125, Hungary}
\altaffiltext{4}{Department of Physics \& Astronomy, San Francisco
	State University, San Francisco, CA 94132, USA}
\altaffiltext{5}{Wise Observatory, Tel Aviv University, Tel Aviv,
    Israel 69978}
\altaffiltext{6}{Department of Terrestrial Magnetism, Carnegie  
	Institute of Washington DC, 5241 Broad Branch Rd.~NW, Washington
	DC, USA 20015-1305}
\altaffiltext{7}{INAF - Osservatorio Astronomico di Torino, 
	Strada Osservatorio 20, 10025, Pino Torinese, Italy}
\altaffiltext{8}{Department of Astronomy,
	E\"otv\"os Lor\'and University, Pf.~32, H-1518 Budapest, Hungary.}
\altaffiltext{9}{Department of Astronomy, University of California,
	Berkeley, CA 94720, USA}
\altaffiltext{10}{Department of Physics, and Kavli Institute for Astrophysics 
	and Space Research, MIT, Cambridge, MA 02139, USA}
\altaffiltext{11}{Hungarian Astronomical Association, 1461 Budapest, 
	P.~O.~Box 219, Hungary}
\altaffiltext{$\dagger$}{
	Based in part on observations obtained at the W.~M.~Keck
	Observatory, which is operated by the University of California and
	the California Institute of Technology. Keck time has been in part
	granted by NASA.
}
\setcounter{footnote}{1}

\begin{abstract}
	We report the discovery of a massive ($\mpl = 9.04 \pm
	0.50\,\mjuplong$) planet transiting the bright ($V=8.7$) F8 star
	\hdcur, with an orbital period of $5.63341\pm0.00013$ days and 
	an eccentricity of $e=0.520\pm0.010$.  From the transit light curve
	we determine that the radius of the planet is
	$\rpl = 0.982\pm^{0.038}_{0.105}\,\rjuplong$. \hdcurb\ (also 
	coined HAT-P-2b) has a mass about 9 times the average mass of
	previously-known transiting exoplanets, and a density of $\rhopl =
	11.9\,\gcmc$, greater than that of rocky planets like the Earth. 
	Its mass and radius are marginally consistent with theories of
	structure of massive giant planets composed of pure H and He, and
	may require a large ($\gtrsim 100\mearth$) core to account for. The
	high eccentricity causes a 9-fold variation of insolation of the
	planet between peri- and apastron. Using follow-up photometry, we
	find that the center of transit is
	$T_{mid}=2,\!454,\!212.8559 \pm 0.0007$ (HJD), and the transit 
	duration is $0.177\pm0.002$\,d.
\end{abstract}

\keywords{
	stars: individual: {\mbox HD 147506} \---
	planetary systems: individual: \hdcurb, HAT-P-2b
}

\section{Introduction}
\label{sec:intro}

To date 18 extrasolar planets have been found which transit their
parent stars and thus yield values for their mass and
radius\footnote{Extrasolar Planets Encyclopedia: http://exoplanet.eu}.
Masses range from 0.3\,\mjup\ to about 1.9\,\mjup, and radii from 0.7\,\rjup\
to about 1.4\,\rjup.  The majority fit approximately what one expects
from theory for irradiated gas giant planets \citep[e.g.][and
references therein]{fortney06}, although there are exceptions:
\hd{149026b} has a small radius for its mass \citep{sato05}, implying
that it has a large heavy core \citep[$\sim$70\,\mearth; ][]{laughlin05},
and several (\hd{209458b}, HAT-P-1b, WASP-1) have unexpectedly large
radii for their masses, perhaps suggesting some presently unknown
source of extra internal heating \citep{guillot02,bodenheimer03}. 
The longest period and lowest density
transiting exoplanet (TEP) detected so far is HAT-P-1b with $P=4.46$\,d
\citep{bakos07}. All TEPs have orbits consistent with circular
Keplerian motion.

From existing radial velocity (RV) data, it might be expected that
there are some close-in (semi-major axis $\lesssim0.07$\,AU, or
$P\lesssim10$\,d) giant planets with masses considerably larger than
any of the 18 transiting planets now known. A well-known example,
considering only objects below the Deuterium burning threshold
\citep[$\sim 13\,\mjup$, e.g.][]{burrows97}, is $\tau$~Boo~b, which
was detected from RV variations, and has a minimum mass of
\mplsini=3.9\,\mjup\ and orbits only 0.046 AU from its star
\citep{butler97}.  Another example is HIP14810~b \citep{wright07} with
similar mass, and orbital period of 6.7\,d and semi-major axis of 
0.069\,AU\@. At this orbital distance the {\em a priori} probability of such a
planet transiting its star is about 10\%.  Thus ``super-massive''
planets should sometimes be found transiting their parent stars.  We
report here the detection of the first such TEP, and our determination
of its mass and radius. This is also the longest period TEP, and the
first one to exhibit highly eccentric orbit.

\section{Observations and Analysis}
\label{sec:anal}

\subsection{Detection of the transit in the HATNet data}
\label{sec:dete}

\notetoeditor{This is the intended place of \figr{lc}. We would like to 
typeset it as a double column figure.}

\ifthenelse{\boolean{emulateapj}}{\begin{figure*}[t]}{\begin{figure}[t]}
\ifpdf
\plotone{f1.pdf}
\else
\plotone{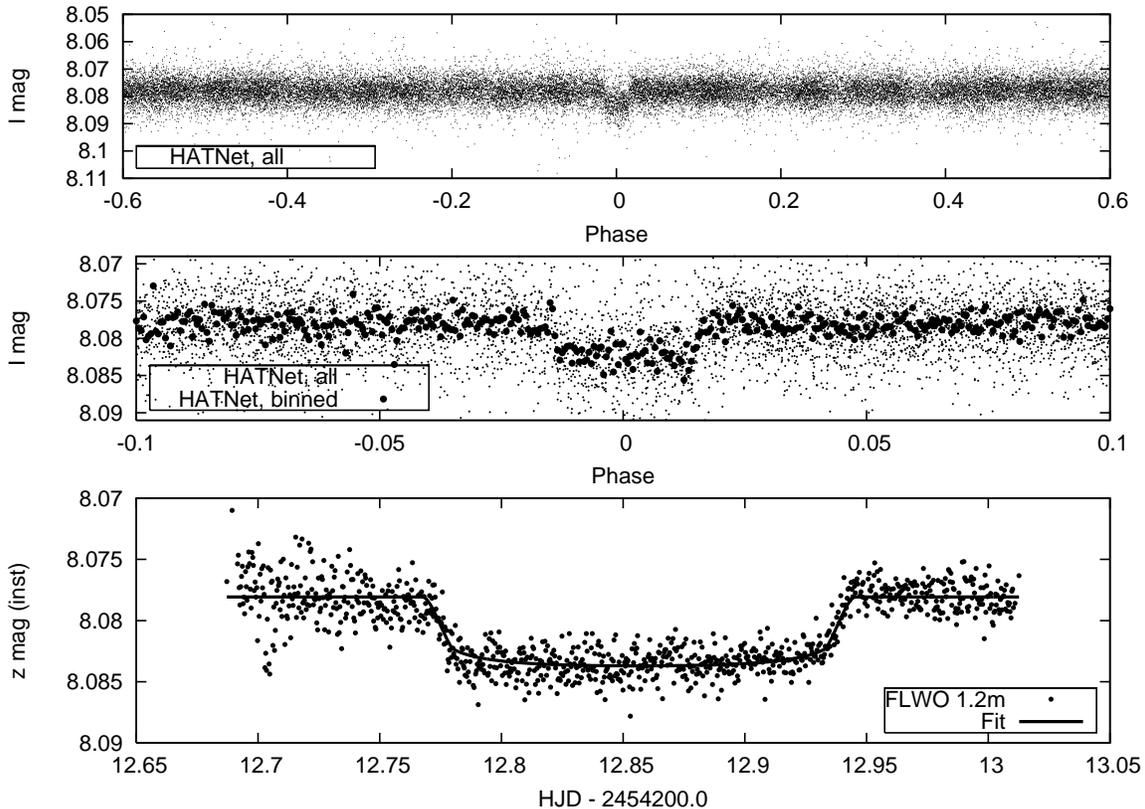}
\fi
\caption{
	The upper panel shows the unbinned HATNet and WHAT joint \lc\ with
	26400 data-points, phased with the $P=5.63341$\,d period. The 5\,mmag
	deep transit is detected with a signal-to-noise of 26. The middle
	panel shows the same HATNet and WHAT data zooming in on the transit
	and binned with a $\phi=0.0005$ bin-size. The lower panel displays
	the Sloan $z$-band photometry taken with the \flwof\ telescope. 
	Over-plotted is our best fit obtained with the \cite{mandel02}
	formalism.
\label{fig:lc}}
\ifthenelse{\boolean{emulateapj}}{\end{figure*}}{\end{figure}}

\hd{147506} is an F8 star with visual magnitude 8.7 and Hipparcos
parallax $7.39\pm0.88$\,mas \citep{perryman97}.  It was initially
identified as a transit candidate in our internally labeled field G193
in the data obtained by HATNet's\footnote{http://www.hatnet.hu}
\citep{bakos02,bakos04} HAT-6 telescope at the Fred Lawrence Whipple
Observatory (FLWO) of the Smithsonian Astrophysical Observatory (SAO).
The detection of a $\sim$5\,mmag transit with a 5.63\,d period in the
\lc\ consisting of $\sim$7000 data-points (with a 5.5\,min cadence) was
marginal. Fortunately the star was in the overlapping corner with
another field (G192) that has been jointly observed by HATNet's HAT-9
telescope at the Submillimeter Array (SMA) site atop Mauna Kea, Hawaii,
and by the Wise HAT telescope \citep[WHAT, Wise-Observatory, Israel;
][]{shporer05}, for an extended period that yielded $\sim$6700 and
$\sim$3900 additional data-points, respectively. The transit was
independently detected and confirmed with these data-sets.  By chance
the candidate is in yet another joint field (G149) of HATNet (HAT-7 at
FLWO) and WHAT, contributing $\sim$6200 and $\sim$2200 additional data
points, respectively. Altogether this resulted in a \lc\ with
exceptional time-coverage (570 days), an unprecedented number of
data-points (26461 measurements at 5.5\,min cadence), and an rms of 
5\,mmag. It is noteworthy that the network coverage by WHAT (longitude 35
E), HATNet at FLWO (111 W) and HATNet at Hawaii (155 W) played an
important role in detecting such a long period and shallow transit.
Data were reduced using astrometry from \citet{pal06}, and with a
highly fine-tuned aperture photometry. We applied our external
parameter decorrelation (EPD) technique on the \lcs, whereby deviations
from the median were cross-correlated with a number of ``external
parameters'', such as the $X$ and $Y$ sub-pixel position, hour-angle,
and zenith distance. We have also applied the Trend Filtering Algorithm
\citep[TFA;][hereafter \citetalias{kovacs05}]{kovacs05} along with the 
Box Least Squares \citep[BLS;][]{kovacs02} transit-search algorithm in
our analysis.  TFA and BLS were combined in signal-reconstruction mode,
assuming general signal shape, as described in \citetalias{kovacs05}.
The detection of this relatively shallow transit is a good
demonstration of the strengths of TFA\@. The upper panel of \figr{lc}
shows the unbinned \lc\ with all 26400 data points, whereas the middle
panel displays the transit binned to 1/2000 of the period (4 minutes).
We note that due to the large amount of data, the binned \lc\ is of
similar precision as a single-transit observation by a 1m-class
telescope.

After several failed attempts (due to bad weather and instrumental
failure) to carry out high-precision photometric follow-up observations
from FLWO, Wise Observatory, Konkoly Observatory, and the Clay Center
(Boston), we finally succeeded in observing a full transit using the
KeplerCam detector on the \flwof\ telescope \citep[see][]{holman07} on
UT 2007 April 22. The Sloan $z$-band light curve is shown in the lower
panel of \figr{lc}. From the combined HATNet and KeplerCam photometry,
spanning a baseline of 839 days, we derive a period of
$5.63341\pm0.00013$\,d and an epoch of mid-transit of 
$T_{mid}=2,\!454,\!212.8559 \pm 0.0007$\,d (HJD). From the \flwof\ data
alone (and the analytic \lc\ fit as described later), the length of
transit is $0.177\pm0.002$\,d (4 hours, 15 minutes), the length of
ingress is $0.012\pm0.002$\,d (17.5 minutes), and the depth (at the
middle of the transit) is 0.0052\,mag.

\subsection{Early spectroscopy follow-up}

Initial follow-up observations were made with the CfA Digital
Speedometer \citep[DS;][]{latham92} in order to characterize the host
star and to reject obvious astrophysical false-positive scenarios that
mimic planetary transits. These observations yielded values of
$\teff=6250\,K$, $\logg=4.0$ and $\vsini=22\,\kms$, corresponding to a
moderately-rotating main sequence F star.  The radial velocity (RV)
measurements showed an rms residual of $\sim$0.82\,\kms, slightly larger
than the nominal DS precision for a star with this rotation, and
suggested that they may be variable. With a few dozen additional DS
observations, it was found that the RV appeared periodic with
$P\approx5.63$\,d, semi-amplitude $\sim$1\,\kms, and phasing in agreement
with predictions from the HATNet+WHAT light curve. This gave strong
evidence that there really was an RV signal resulting from Keplerian
motion, although the precision was insufficient to establish the orbit
with confidence. Altogether we collected 53 individual spectra spanning
a time-base of more than a year (\tabr{rv}).

\notetoeditor{This is the intended place of \tabr{rv}. In the journal
	paper we give only a short extract of the real table. In the
	electronic edition the input file is \verb#tab_rv_data.tex#}
\begin{deluxetable}{lrrr}
\tabletypesize{\scriptsize}
\tablewidth{0pt}
\tablecaption{
	\label{tab:rv}
	Radial Velocities for \hdcur.
}
\tablewidth{0pt}
\tablehead{
	\colhead{BJD $- 2,\!400,\!000$} &
	\colhead{RV\tablenotemark{a}} &
	\colhead{Uncert.} &
	\colhead{Observatory\tablenotemark{b}} \\
	\colhead{(days)} &
	\colhead{(\ms)} &
	\colhead{(\ms)} &
}
\startdata
$53981.7775$ &	$-556.0$ &	$8.4$ &	Keck\\
$53982.8717$ &	$-864.1$ &	$8.5$ &	Keck\\
$53983.8148$ &	$-62.9$ &	$8.8$ &	Keck\\
$53984.8950$ &	$280.6$ &	$8.6$ &	Keck\\
$54023.6915$ &	$157.8$ &	$9.9$ &	Keck\\
$54186.9982$ &	$120.2$ &	$5.5$ &	Keck\\
$54187.1041$ &	$104.6$ &	$5.7$ &	Keck\\
$54187.1599$ &	$130.1$ &	$5.3$ &	Keck\\
$54188.0169$ &	$168.5$ &	$5.3$ &	Keck\\
$54188.1596$ &	$198.2$ &	$5.5$ &	Keck\\
$54189.0104$ &	$68.9$ &	$5.7$ &	Keck\\
$54189.0889$ &	$69.7$ &	$6.2$ &	Keck\\
$54189.1577$ &	$25.2$ &	$6.1$ &	Keck\\
$54168.9679$ &	$-152.7$ &	$42.1$ &	Lick\\
$54169.9519$ &	$542.4$ &	$41.3$ &	Lick\\
$54170.8619$ &	$556.8$ &	$42.6$ &	Lick\\
$54171.0365$ &	$719.1$ &	$49.6$ &	Lick\\
$54218.8081$ &	$-1165.2$ &	$88.3$ &	Lick\\
$54218.9856$ &	$-1492.6$ &	$90.8$ &	Lick\\
$54219.9373$ &	$-28.2$ &	$43.9$ &	Lick\\
$54219.9600$ &	$-14.8$ &	$43.9$ &	Lick\\
$54220.9641$ &	$451.6$ &	$38.4$ &	Lick\\
$54220.9934$ &	$590.7$ &	$37.1$ &	Lick\\
\enddata
\tablenotetext{a}{The RVs include the barycentric correction.}
\tablenotetext{b}{Only the Keck and Lick data-points are
	shown here. Consult the electronic edition for a full data-set
	that includes the CfA DS measurements.}
\end{deluxetable}

\subsection{High-precision spectroscopy follow-up}

In order to confirm or refute the planetary nature of the transiting
object, we pursued follow-up observations with the HIRES instrument
\citep{vogt94} on the W.~M.~Keck telescope and with the Hamilton
Echelle spectrograph at the Lick Observatory \citep{vogt87}. The
spectrometer slit used at Keck is $0\farcs 86$, yielding a resolving
power of about $55,\!000$ with a spectral coverage between about 3200
and 8800\,\AA\@. The Hamilton Echelle spectrograph at Lick has a similar
resolution of about $50,\!000$. These spectra were used to i) more
fully characterize the stellar properties of the system, ii) to obtain
a radial velocity orbit, and to iii) check for spectral line bisector
variations that may be indicative of a blend. We gathered 13 spectra at
Keck (plus an iodine-free template) spanning 207 days, and 10 spectra
at Lick (plus template) spanning 50 days. The radial velocities
measured from these spectra are shown in \tabr{rv}, along with those
from the CfA DS.

\section{Stellar parameters}
\label{sec:stelpar}

A spectral synthesis modeling of the iodine-free Keck template spectrum
was carried out using the SME software \citep{valenti96}, with the
wavelength ranges and atomic line data described by \citet{valenti05}. 
Results are shown in \tabr{stelpar}.  The values obtained for the
effective temperature (\teff), surface gravity (\logg), and projected
rotational velocity ($\vsini$) are consistent with those found from the
CfA DS spectra. As a check on \teff, we collected all available
photometry for \hdcur\ in the Johnson, Cousins, 2MASS, and Tycho
systems, and applied a number of color-temperature calibrations
\citep{ramirez05, masana06, casagrande06} using 7 different color
indices. These resulted in an average temperature of $\sim$6400$ \pm
100$\,K, somewhat higher than the spectroscopic value but consistent
within the errors.

Based on the Hipparcos parallax ($\pi = 7.39\pm0.88$\,mas), the apparent
magnitude $V = 8.71 \pm 0.01$ \citep{droege06}, the SME temperature,
and a bolometric correction of $BC_V = -0.011 \pm 0.011$\,mag
\citep{flower96}, application of the Stefan-Boltzmann law yields a
stellar radius of $\rstar = 1.84\pm 0.24\,\rsun$.

\notetoeditor{This is the intended place of \tabr{stelpar}}
\newboolean{alltable}
\setboolean{alltable}{false}
\begin{deluxetable}{lll}
\tabletypesize{\scriptsize}
\tablecaption{
	Summary of stellar parameters for \hdcur.
\label{tab:stelpar}}
\tablehead{
	\colhead{Parameter} &
	\colhead{Value} &
	\colhead{Source}
}
\startdata
\ifthenelse{\boolean{alltable}}{
	\teff (K)			&	$6290\pm110$	& SME\\
	\logg 				&	$4.22\pm0.14$	& SME\\
	\vsini (\kms)		&	$19.8\pm1.6$	& SME\\
	\feh (dex)			&	$+0.12\pm0.08$	& SME\\\hline
	Distance (pc)		&	$135\pm16$				& HIP\\
	Distance (pc)		&	$110\pm15$					& Y$^2$ isochrones, $a/\rstar$ constraint\\
	\logg				&	$4.214\pm^{0.085}_{0.015}$	& Y$^2$ isochrones, $a/\rstar$ constraint\\
	Mass (\msun)		&	$1.298\pm^{0.062}_{0.098}$	& Y$^2$ isochrones, $a/\rstar$ constraint\\
	Radius (\rsun)		&	$1.474\pm^{0.042}_{0.167}$	& Y$^2$ isochrones, $a/\rstar$ constraint\\
	$\loglstar$ (\lsun)	&	$0.485 \pm^{0.052}_{0.134}$	& Y$^2$ isochrones, $a/\rstar$ constraint\\
	Age (Gyr)			&	$2.7\pm^{0.8}_{1.4}$		& Y$^2$ isochrones, $a/\rstar$ constraint\\\hline
	$a/\rstar$			&	$9.954\pm^{1.914}_{1.645}$	& Y$^2$ isochrones, \logg\ constraint\\
	Mass (\msun)		&	$1.292\pm^{0.170}_{0.116}$	& Y$^2$ isochrones, \logg\ constraint\\
	Radius (\rsun)		&	$1.460\pm^{0.361}_{0.273}$	& Y$^2$ isochrones, \logg\ constraint\\
	Age (Gyr)			&	$2.60\pm^{0.8}_{2.5}$		& Y$^2$ isochrones, \logg\ constraint\\\hline
	$a/\rstar$			&	$8.090\pm^{1.354}_{1.187}$	& Y$^2$ isochrones, $M_V$ constraint\\
	Mass (\msun)		&	$1.418\pm^{0.100}_{0.118}$	& Y$^2$ isochrones, $M_V$ constraint\\
	Radius (\rsun)		&	$1.853\pm^{0.314}_{0.280}$	& Y$^2$ isochrones, $M_V$ constraint\\
	Age (Gyr)			&	$2.70\pm^{1.4}_{0.6}$		& Y$^2$ isochrones, $M_V$ constraint\\
}{
	\teff (K)			&	$6290\pm110$	& SME\\
	\logg 				&	$4.22\pm0.14$	& SME\\
	\vsini (\kms)		&	$19.8\pm1.6$	& SME\\
	\feh (dex)			&	$+0.12\pm0.08$	& SME\\
	Distance (pc)		&	$135\pm16$				& HIP\\
	Distance (pc)		&	$110\pm15$				& Y$^2$ isochrones, $a/\rstar$ constraint\\
	\logg				&	$4.214\pm^{0.085}_{0.015}$	& Y$^2$ isochrones, $a/\rstar$ constraint\\
	Mass (\msun)		&	$1.298\pm^{0.062}_{0.098}$	& Y$^2$ isochrones, $a/\rstar$ constraint\\
	Radius (\rsun)		&	$1.474\pm^{0.042}_{0.167}$	& Y$^2$ isochrones, $a/\rstar$ constraint\\
	$\loglstar$ (\lsun)	&	$0.485 \pm^{0.052}_{0.134}$	& Y$^2$ isochrones, $a/\rstar$ constraint\\
	$M_V$				&	$3.54\pm^{0.36}_{0.15}$		& Y$^2$ isochrones, $a/\rstar$ constraint\\
	Age (Gyr)			&	$2.6\pm^{0.8}_{1.4}$		& Y$^2$ isochrones, $a/\rstar$ constraint\\
}
\enddata
\end{deluxetable}

A more sophisticated approach to determine the stellar parameters uses
stellar evolution models along with the observational constraints from
spectroscopy. For this we used the Y$^2$ models by \citet{yi01} and
\citet{demarque04}, and explored a wide range of ages to find all
models consistent with $T_{\rm eff}$, $M_V$, and [Fe/H] within the
observational errors. Here $M_V = 3.05 \pm 0.26$ is the absolute visual
magnitude, as calculated from $V$ and the Hipparcos parallax. This
resulted in a mass and radius for the star of 
$\mstar = 1.42\pm^{0.10}_{0.12}\,\msun$ and 
$\rstar = 1.85\pm^{0.31}_{0.28}\,\rsun$,
and a best-fit age of $2.7^{+1.4}_{-0.6}$\,Gyr.
Other methods that rely on the Hipparcos parallax, such as the
Padova\footnote{http://pleiadi.pd.astro.it/} stellar model grids
\citep{girardi02}, consistently yielded a stellar mass of
$\sim1.4\,\msun$ and stellar radius $\sim1.8\,\rsun$.

If we do {\em not} rely on the Hipparcos parallax, and use \logg\ as a
proxy for luminosity (instead of $M_V$), then the Y$^2$ stellar evolution
models yield a smaller stellar mass of
$\mstar = 1.29\pm^{0.17}_{0.12}\,\msun$, 
radius of
$\rstar = 1.46\pm^{0.36}_{0.27}\,\rsun$, and best fit age of
$2.6^{+0.8}_{-2.5}$\,Gyr.
The surface gravity is a sensitive measure of the degree of evolution
of the star, as is luminosity, and therefore has a very strong
influence on the radius. However, $\log g$ is a notoriously difficult
quantity to measure spectroscopically and is often strongly correlated
with other spectroscopic parameters.

It has been pointed out by \citet{sozzetti07} that the normalized
separation $a/\rstar$ can provide a much better constraint for stellar
parameter determination than \logg. The $a/\rstar$ quantity can be
determined directly from the photometric observations, without
additional assumptions, and it is related to the density of the central
star. As discussed later in \secr{planpar}, an analytic fit to the
\flwof\ \lc, taking into account an eccentric orbit, 
yielded $a/\rstar = 9.77^{+1.10}_{-0.02}$. Using this as a
constraint, along with \teff\ and \feh, we obtained 
$\mstar = 1.30\pm^{0.06}_{0.10}\,\msun$, 
$\rstar = 1.47\pm^{0.04}_{0.17}\,\rsun$ and age of
$2.6^{+0.8}_{-1.4}$\,Gyr. 
The $\logg = 4.214\pm^{0.085}_{0.015}$ derived this way is consistent
with former value from SME.

As seen from the above discussion, there is an inconsistency between
stellar parameters depending on whether the Hipparcos parallax is
employed or not. Methods relying on the parallax (Stefan-Boltzmann law,
stellar evolution models with $M_V$ constraint, etc) prefer a larger
mass and radius ($\sim1.4\,\msun$, $\sim1.8\,\rsun$, respectively), whereas
methods that do not rely on the parallax (stellar evolution models with
\logg\ or $a/\rstar$ constraint) point to smaller mass and radius
($\sim1.3\,\msun$, $\sim1.46\,\rsun$, respectively). 
We have chosen to rely on the $a/\rstar$ method, which yields
considerably smaller uncertainties and a calculated transit duration
that matches the observations. Additionally, it implies an angular
diameter for the star ($\phi = 0.127^{+0.021}_{-0.014}$ mas) that is in
agreement with the more direct estimate of $\phi = 0.117 \pm 0.001$ mas
from the near-infrared surface-brightness relation by
\citet{kervella04}. The later estimate depends only on the measured
$V-K_s$ color and apparent $K_s$ magnitude (ignoring extinction) from
2MASS \citep{skrutskie06}, properly converted to the homogenized
Bessell \& Brett system for this application
\citep[following][]{carpenter01}. We note that our results from the
$a/\rstar$ method imply a somewhat smaller distance to \hd{147506} than
the one based on the Hipparcos parallax.  The final adopted stellar
parameters are listed in \tabr{stelpar}.

\subsection{Stellar jitter}

Stars with significant rotation are known to exhibit excess scatter
(``jitter'') in their radial velocities \citep[e.g.,][and references
therein]{wright05}, due to enhanced chromospheric activity and the
associated surface inhomogeneities (spottedness). This jitter is in
addition to the internal errors in the measured velocities, and could
potentially be significant in our case. We note that after
pre-whitening the \lc\ with the transit component, we found no
significant sinusoidal signal above 0.3\,mmag amplitude. From this we
conclude that there is no very significant spot activity on the star
(in the observed 500 day window). In order to estimate the level of
chromospheric activity in the star, we have derived an activity index
from the \ion{Ca}{2} H and K lines in our Keck spectra of $\logrhk =
-4.72\pm0.05$. For this value the calibration by \cite{wright05}
predicts velocity jitter ranging from 8 to 16\,\ms. An earlier
calibration by \citet{saar98}, parametrized in terms of the projected
rotational velocity, predicts a jitter level of up to 50\,\ms\ for our
measured $\vsini$ of $20\,\kms$. A different calibration by the same
authors in terms of \rhk\ gives 20\,\ms.  An additional way to estimate
the jitter is to compare \hdcur\ to stars of the Lick Planet Search
program \citep{cumming99} that have similar properties ($0.4<B-V<0.5$,
$\vsini > 15\,\kms$). There are four such stars (J.~Johnson, private
communication), and their average jitter is $45\,\ms$. A more direct
measure for the particular case of \hdcur\ may be obtained from the
multiple exposures we collected during a 3-night Keck run in 2007
March. Ignoring the small velocity variations due to orbital motion
during any given night, the overall scatter of these 8 exposures
relative to the nightly means is $\sim$20\,\ms. This may be taken as an
estimate of the jitter on short timescales, although it could be
somewhat larger over the entire span of our observations. Altogether,
it is reasonable to expect the jitter to be at least 10\,\ms, and
possibly around 30\---50\,\ms\ for this star.

\notetoeditor{This is the intended place of \figr{rv}. We would like to 
typeset it as a single column figure.}
\ifthenelse{\boolean{emulateapj}}{\begin{figure}[t]}{\begin{figure}[t]}
\ifpdf
\plotone{f2.pdf}
\else
\plotone{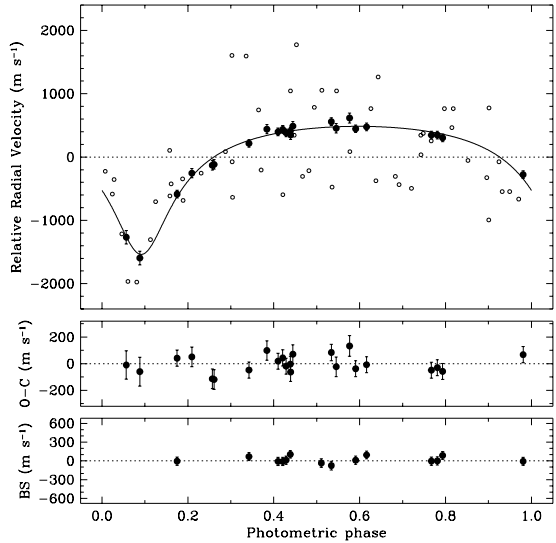}
\fi
\caption{
	The upper panel shows the RV measurements phased with the period of
	$P=5.63341\,d$. The zero-point in phase corresponds to the epoch of
	mid-transit. Large filled circles indicate Keck and Lick points. 
	Small open circles denote CfA DS data (not used for the fit). 
	Overlaid is the fit that was based only on the Keck and Lick data
	assuming 60\,\ms\ stellar jitter. The middle panel shows the
	residuals from the fit. The lower panel exhibits the line bisector
	spans on the same scale as the upper panel. No variation in the
	line bisectors is seen concomitant with that in the RVs,
	essentially confirming the planetary nature of the transiting
	object.
\label{fig:rv}}
\ifthenelse{\boolean{emulateapj}}{\end{figure}}{\end{figure}}

\section{Spectroscopic orbital solution}

We have three velocity data sets available for analysis: 13 relative
radial velocity measurements from Keck, 10 from Lick, and 53
measurements from the CfA DS, which are nominally on an absolute scale
(\tabr{rv}).  Given the potential effect of stellar jitter, we
performed weighted Keplerian orbital solutions for a range of jitter
values from 10 to 80\,\ms\ with 10\,\ms\ steps. These jitter values were
added in quadrature to all individual internal errors. We performed
separate fits for the star orbited by a single planet, both with and
without the CfA DS measurements, since these have errors
($\sim$600\,\ms) significantly larger than Keck (5--9\,\ms) or Lick
(40--90\,\ms). In all of these solutions we held the period and transit
epoch fixed at the photometric values given earlier. The parameters
adjusted are the velocity semi-amplitude $K$, the eccentricity $e$, the
longitude of periastron $\omega$, the center-of-mass velocity for the
Keck relative velocities $\gamma$, and offsets $\Delta v_{KL}$ between
Keck and Lick and $\Delta v_{KC}$ between Keck and CfA DS\@. The fitted
parameters were found to be fairly insensitive to the level of jitter
assumed. However, only for a jitter of $\sim$60\,\ms\ (or $\sim$70\,\ms\
when the CfA DS data are included) did the $\chi^2$ approach values
expected from the number of degrees of freedom. There are thus two
possible conclusions: if we accept that \hdcur\ has stellar jitter at
the 60\,\ms\ level, then a single-planet solution such as ours
adequately describes our observations. If, on the other hand, the true
jitter is much smaller ($\lesssim 20\,\ms$), then the extra scatter
requires further explanation (see below).  Our adopted orbital
parameters for the simplest single-planet Keplerian solution are based
only on the more precise Keck and Lick data, and assume the jitter is
60\,\ms\ (\tabr{orb}). The orbital fit is shown graphically in the upper
panel of \figr{rv}. In this figure, the zero-point of phase is chosen
to occur at the epoch of mid-transit, $T_{mid}=2,\!454,\!212.8559$
(HJD). The most significant results are the large eccentricity ($e =
0.520 \pm 0.010$), and the large velocity semi-amplitude
($K=1011\pm38\,\ms$, indicating a very massive companion). As we show in
the next section (\secr{blend}), the companion is a planet,
i.e.~\hdcurb, which we hereafter refer to as HAT-P-2b.

As a consistency check we also fitted the orbits by fixing only the
period, and leaving the transit epoch as a free parameter. We found
that for all values of the stellar jitter the predicted time of
transit as derived from the RV fit was consistent with the photometric
ephemeris within the uncertainties. We also found that in these
fits the orbital parameters were insensitive to the level of
jitter and to whether or not the CfA DS data were included. The
eccentricity values ranged from 0.51 to 0.53.

\subsection{Solutions involving two planets}

If we assume that the true stellar jitter is small, then the excess
scatter in the RV fit could be explained by a third body in the system,
i.e., a hypothetical HAT-P-2c. In addition, such a body could provide a
natural dynamical explanation for the large eccentricity of HAT-P-2b at
this relatively short period orbit. Preliminary two-planet orbital fits
using all the data yielded solutions only significant at the 2-sigma
level, not compelling enough to consider as evidence for such a
configuration. Additional RV measurements are needed to firmly
establish or refute the existence of HAT-P-2c.

We also exploited the fact that the HATNet \lc\ has a unique time
coverage and precision, and searched for signs of a second transit that
might be due to another orbiting body around the host star. Successive
box-prewhitening based on the BLS spectrum and assuming
trapezoidal-shape transits revealed no secondary transit deeper than
the 0.1\% level and period $\lesssim 10$ days.

\section{Excluding blend scenarios}
\label{sec:blend}

As an initial test to explore the possibility that the photometric
signal we detect is a false positive (blend) due to contamination from
an unresolved eclipsing binary, we modeled the light curve assuming
there are three coeval stars in the system, as described by
\citet{torres04}. We were indeed able to reproduce the observed light
curve with a configuration in which the brighter object is accompanied
by a slightly smaller F star which is in turn being eclipsed by a
late-type M dwarf.  However, the predicted relative brightness of the
two brighter objects at optical wavelengths would be $\sim0.58$, and
this would have been easily detected in our spectra. This configuration
can thus be ruled out.

The reality of the velocity variations was tested by carefully
examining the spectral line bisectors of the star in our more numerous
Keck spectra. If the velocity changes measured are due only to
distortions in the line profiles arising from contamination of the
spectrum by the presence of a binary with a period of 5.63 days, we
would expect the bisector spans (which measure line asymmetry) to vary
with this period and with an amplitude similar to the velocities
\citep[see, e.g.,][]{queloz01,torres05}. The bisector spans were
computed from the cross-correlation function averaged over 15 spectral
orders blueward of 5000\,\AA\ and unaffected by the iodine lines, which
is representative of the average spectral line profile of the star. The
cross correlations were performed against a synthetic spectrum matching
the effective temperature, surface gravity, and rotational broadening
of the star as determined from the SME analysis. As shown in \figr{rv},
while the measured velocities exhibit significant variation as a
function of phase (upper panel), the bisector spans are essentially
constant within the errors (lower panel).  Therefore, this analysis
rules out a blend scenario, and confirms that the orbiting body is
indeed a planet.

\section{Planetary parameters}
\label{sec:planpar}

For a precise determination of the physical properties of HAT-P-2b we
have modeled the \flwof\ Sloan $z$-band photometric data shown in
\figr{lc}. The model is an eccentric Keplerian orbit of a star and
planet, thus accounting for the nonuniform speed of the planet and the
reflex motion of the star. Outside of transits, the model flux is
unity. During transits, the model flux is computed using the formalism
of \citet{mandel02}, which provides an analytic approximation of the
flux of a limb-darkened star that is being eclipsed. The free
parameters were the mid-transit time $T_{mid}$, the radius ratio
$R_p/R_\star$, the orbital inclination $i$, and the scale parameter
$a/\rstar$, where $a$ is the semimajor axis of the relative orbit. The
latter parameter is determined by the time scales of the transit (the
total duration and the partial-transit duration), and is related to the
mean density of the star (see \secr{stelpar}). The orbital period,
eccentricity, and argument of pericenter were fixed at the values
determined previously by fitting the radial velocity data. The limb
darkening law was assumed to be quadratic, with coefficients taken from
\citet{claret04}.

To solve for the parameters and their uncertainties, we used a Markov
Chain Monte Carlo algorithm that has been used extensively for modeling
other transits \citep[see, e.g.][]{winn07,holman07}. This algorithm
determines the {\em a posteriori} probability distribution for each
parameter, assuming independent (``white'') Gaussian noise in the
photometric data. However, we found that there are indeed correlated
errors. Following Gillon et al.~(2006), we estimated the red noise
$\sigma_r$ via the equation
\begin{equation}
\sigma_r^2 = \frac{\sigma_N^2 - \sigma_1^2/N}{1-1/N},
\end{equation}
where $\sigma_1$ is the standard deviation of the out-of-transit flux
of the original (unbinned) light curve, $\sigma_N$ is the standard
deviation of the light curve after binning into groups of $N$ data
points, and $N=40$ corresponds to a binning duration of 20 minutes,
which is the ingress/egress time scale that is critical for parameter
estimation. With white noise only, $\sigma_N=\sigma_1/\sqrt{N}$ and
$\sigma_r=0$. We added $\sigma_r$ in quadrature to the error bar of
each point, effectively inflating the error bars by a factor of 1.25.

The result for the radius ratio is $\rpl/\rstar=0.0684\pm0.0009$, and
for the scale parameter $a/\rstar = 9.77^{+1.10}_{-0.02}$.  The {\em a
posteriori} distribution for $a/\rstar$ is very asymmetric because the
transit is consistent with being equatorial: $i>84\fdg 6$ with 95\%
confidence. We confirmed that these uncertainties are dominated by the
photometric errors, rather than by the covariances with the orbital
parameters $e$, $\omega$, and $P$, and hence we were justified in
fixing those orbital parameters at constant values. Based on the
inclination, the mass of the star (\tabr{stelpar}) and the orbital
parameters (\tabr{orb}), the planet mass is then $9.04 \pm 0.50\,\mjup$.
Based on the radius of the star (\tabr{stelpar}) and the above
$\rpl/\rstar$ determination, the radius of the planet is $\rpl =
0.982\pm^{0.038}_{0.105}\,\rjup$. These properties are summarized in
\tabr{orb}.

\notetoeditor{This is the intended place of \tabr{orb}}
\begin{deluxetable}{lr}
\tablecaption{
	\label{tab:orb}
	Orbital fit and planetary parameters for the HAT-P-2 system.
}
\tablehead{
	\colhead{Parameter} &
	\colhead{Value}
} 
\startdata
Period (d)\tablenotemark{a}			&	$5.63341\pm0.00013$\\
$T_{mid}$ (HJD)\tablenotemark{a}	&	$2,\!454,\!212.8559\pm0.0007$\\
Transit duration (day)				&	$0.177\pm 0.002$\\
Ingress duration (day)				&	$0.012\pm 0.002$\\\hline
Stellar jitter (\ms)\tablenotemark{b}		&	60\\
$\gamma$ (\ms)\tablenotemark{c}		&	$-278\pm20$\\
$K$ (\ms)							&	$1011\pm38$		\\
$\omega$ (deg)						&	$179.3\pm3.6$\\
$e$									&	$0.520\pm0.010$\\
$T_{peri}$ (HJD)					&	$2,\!454,\!213.369\pm0.041$\\
$\Delta v_{KL}$ (\ms)				&	$-380\pm35$\\
$\Delta v_{KC}$ (\kms)\tablenotemark{d}			&	$19.827\pm0.087$\\
$f(M)$ (\msun)						&	$(376 \pm 42) \times 10^{-9}$\\
$\mpl\sin i$ (\mjup)				&	$7.56 \pm 0.28 ([\mstar + \mpl]/M_{\sun}])^{2/3}$\\
$\astar \sin i$ (km)				&	$(0.0669 \pm 0.0025) \times 10^6$\\\hline
$a_{\rm rel}$ (AU)					&	$0.0677\pm0.0014$		\\
\ipl (deg)							&	$>84.6\arcdeg$ (95\% confidence)\\
\mpl (\mjup)						&   $9.04\pm0.50$		\\
\rpl (\rjup)						&	$0.982\pm^{0.038}_{0.105}$ \\
\rhopl (\gcmc)						&	$11.9\pm^{4.8}_{1.6}$\\
\gpl ($m\,s^{-2}$)					&	$227\pm^{44}_{16}$\\
\enddata
\tablenotetext{a}{Fixed in the orbital fit.}
\tablenotetext{b}{Adopted (see text).}
\tablenotetext{c}{The $\gamma$ velocity is not in an absolute reference frame.}
\tablenotetext{d}{The offset between Keck and CfA DS is given for reference
from a fit that includes all data sets,	but does not affect our solution.}
\end{deluxetable}

\section{Discussion}
\label{sec:disc}

\notetoeditor{This is the intended place of \figr{exomr}. We would like to 
typeset it as a single column figure.}

\ifthenelse{\boolean{emulateapj}}{\begin{figure}[t]}{\begin{figure}[t]}
\ifpdf
\plotone{f3.pdf}
\else
\plotone{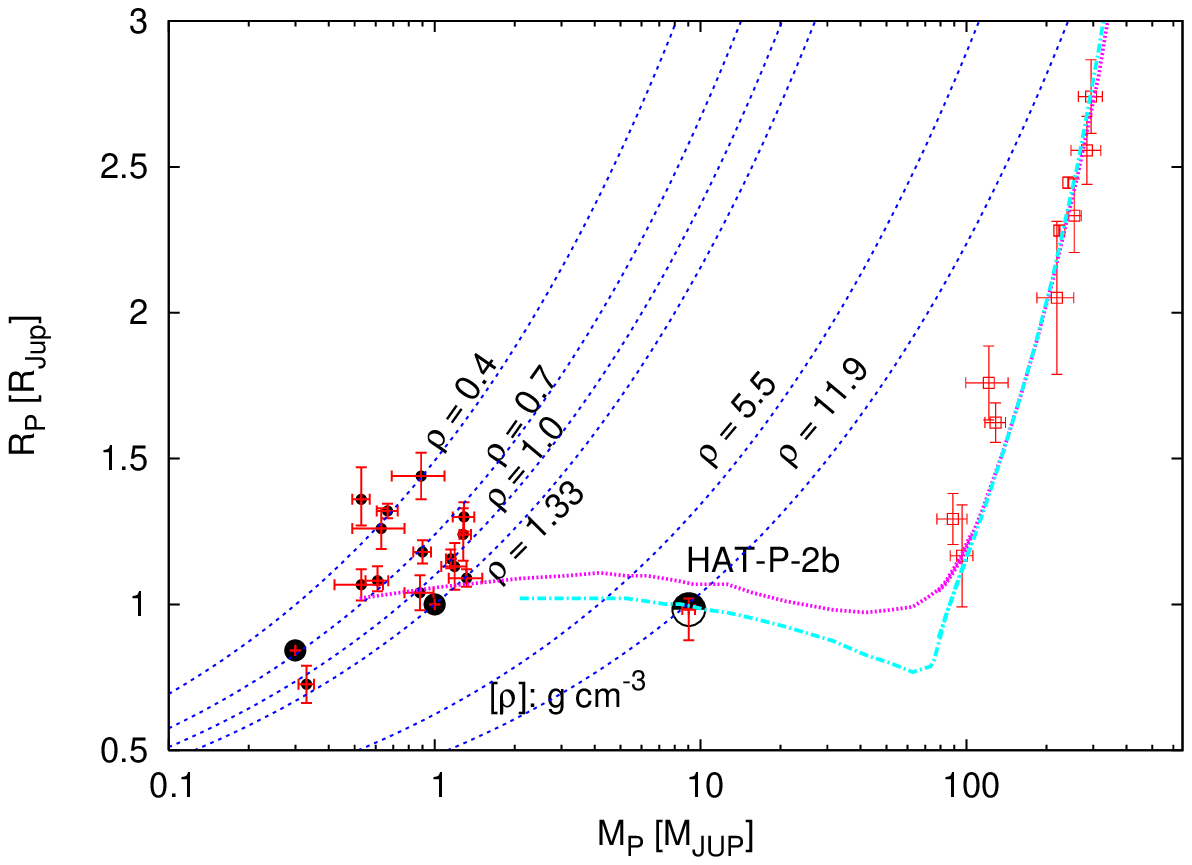}
\fi
\caption{
	The mass--radius diagram of known TEPs
	(from www.exoplanet.eu and references therein), Jupiter and Saturn
	(large filled circles), and low-mass stars from \citet{beatty07}. 
	HAT-P-2b is an intermediate mass object that is still in the
	planetary regime (well below 13\,\mjup). Overlaid are equidensity
	lines (labeled), \citet{baraffe98} (stellar) and
	\citet{baraffe03} (zero insolation planetary) isochrones for ages
	of 0.5\,Gyr (upper, dotted line) and 5\,Gyr (lower dashed-dotted
	line), respectively.
\label{fig:exomr}}
\ifthenelse{\boolean{emulateapj}}{\end{figure}}{\end{figure}}

In comparison with the other 18 previously known transiting exoplanets,
HAT-P-2b is quite remarkable (\figr{exomr}, \figr{Pg}).  Its mass of
$9.04\pm0.50\,\mjup$ is $\sim$5 times greater than any of these 18 other
exoplanets.  Its mean density $\rho = 11.9\pm^{4.8}_{1.6}\,\gcmc$ is
$\sim$9 times that of the densest known exoplanet (OGLE-TR-113b, $\rho
= 1.35\,\gcmc$) and indeed greater than that of the rocky planets of the
Solar System ($\rho$ = 5.5\,\gcmc).  Its surface gravity of
$227\pm^{44}_{16}\,\mss$ is 7 times that of any of the previously known
TEPs, and 30 times that of HAT-P-1b (\figr{Pg}).

We may compare the mass and radius for HAT-P-2b with evolutionary
models, including irradiation, as recently presented by
\citet[][]{fortney06} (hereafter \citetalias{fortney06}). 
Given the inferred stellar luminosity (\tabr{stelpar}), and the
time-integral of the insolation over an entire period (taking into
account the orbital parameters, notably $e$ and $a_{rel}$), the
equivalent semi-major axis $a_{eq}$ for the same amount of irradiation
if the central star were solar is 0.036 AU\@. At that separation,
\citetalias{fortney06} find for a pure hydrogen/helium planet of mass
9\,\mjup\ and age of 4.5\,Gyr a planetary radius about 1.097\,\rjup. A
100\,\mearth\ core has a negligible effect on the radius (yielding
1.068\,\rjup), which is not surprising, since the mass of such a core
is only a few percent of the total mass. For younger ages of 1\,Gyr and
0.3\,Gyr the radii are larger: 1.159\,\rjup\ and 1.22\,\rjup\ for
coreless models, respectively. Our observed radius of $0.982\,\rjup$ is
smaller than any of the above values (4.5, 1, 0.3\,Gyr, with or without
100\,\mearth\ core). Since the $1\sigma$ positive error-bar on our
radius determination is 0.038\,\rjup, the inconsistency is only
marginal. Nevertheless, the observed radius prefers either larger age
or bigger core-size, or both. Given the age of the host star (2.6\,Gyr,
\tabr{stelpar}) the larger age is an unlikely explanation. The required
core-size for this mass and radius according to \citetalias{fortney06} 
would be 300\,\mearth, which amount of icy and rocky material may be 
hard to account for.

\figr{exomr} also shows a theoretical mass-radius relation for objects
ranging from gas giant planets to stars \citep{baraffe98,baraffe03}.
Note that HAT-P-2b falls on the relation connecting giant
planets to brown dwarfs to stars.  It thus appears to be intermediate
in its properties between Jupiter-like planets and more massive objects
like brown dwarfs or even low mass stars. According to theories, stars
with mass $\gtrsim 0.2\,\msun$ have a core, where internal pressure is
dominated by classical gas (ions and electrons), and the $R \propto M$
radius--mass relation holds in hydrostatic equilibrium \citep[for a
review and details on the following relations see e.g.][]{chabrier00}.
Below $\sim 0.075\,\msun\ (80\,\mjup)$ mass, however, the equation of state
in the core becomes dominated by degenerate electron gas ($R \propto
M^{-1/3}$ for full degeneracy), yielding an expected minimum in the
mass--radius relationship (around 73\,\mjup). Below this mass, the
partial degeneracy of the object and the classical ($R\propto M^{1/3}$)
Coulomb pressure together yield an almost constant radius ($R\propto
M^{-1/8}$). HAT-P-2b is a demonstration of this well known
phenomenon.  (The approximate relation breaks below $M\sim4\,\mjup$,
where the degeneracy saturates, and a classical mass--radius behaviour
is recovered).

\notetoeditor{This is the intended place of \figr{Pg}. We would like to 
typeset it as a single column figure.}

\ifthenelse{\boolean{emulateapj}}{\begin{figure}[t]}{\begin{figure}[t]}
\ifpdf
\plotone{f4.pdf}
\else
\plotone{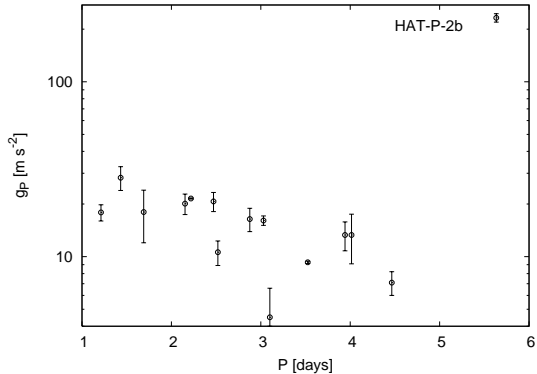}
\fi
\caption{
	The surface gravity of TEPs as a function of orbital period.  Data
	taken from \citet{southworth07} with the exception of HAT-P-2b.
	This object is clearly not obeying the suspected correlation of
	$\gpl$ and $P$ for ``Jupiter-mass'' objects.
\label{fig:Pg}}
\ifthenelse{\boolean{emulateapj}}{\end{figure}}{\end{figure}}

Compared to the other 18 known transiting planets, HAT-P-2b is also
unique in having an orbit with remarkably high eccentricity. 
The primary question is how such an eccentricity was created in the
first place. One possible explanation could be that the planet was
scattered inward from a larger orbit, acquiring a high eccentricity in
the process \citep{ford07,chatterjee07}. If so, then the scattering 
event might
have caused its new orbital plane to be inclined relative to the plane
of the original disk, and hence out of the equatorial plane of the
parent star \citep[e.g.][]{fabrycky07}.  
This angle between these two planes should be readily
measurable from the Rossiter-McLaughlin effect \citep{winn05}. Indeed,
the star \hdcur\ is an ideal subject for studying this effect, because
its rapid rotation should lead to a relatively large Rossiter-McLaughlin 
signal.

There are a number of other interesting issues related to the high
eccentricity of HAT-P-2b.  During its 5.63 day orbit, the insolation
reaching the planet's surface varies by a factor of 9. Assuming an
albedo of 0.1 \citep{rowe06} and complete redistribution of insolation
energy over the surface of the planet, the equilibrium temperature
varies from about 2150\,K at periastron to 1240\,K at apastron.  This
would have a major influence on atmospheric dynamics and
photochemistry.

It is interesting to compare the properties of the HAT-P-2 system with
the $\tau$~Boo system, which \--- as already noted \--- harbors a
close-in planet with minimum mass $\mplsini = 3.9\,\mjup$.  Similarities
of the two parent stars include the nearly identical masses, effective
temperature, and the rapid rotation, although $\tau$~Boo, with
$\feh = +0.28$, is somewhat more metal rich than \hdcur, with $\feh =
+0.12$.  A striking difference is that, while the orbital eccentricity
of HAT-P-2b is 0.5, the eccentricity of $\tau$~Boo~b is not measurably
different from zero. However, $\tau$~Boo~b's orbital period, 3.3 days,
is almost half that of HAT-P-2b. A large fraction of close-in planets
with $5<P<10$ days have significant eccentricities ($0.1<e<0.3$)
although not as large as HAT-P-2b. For discussion on the eccentricity
distribution see \citet{juric07}. As circularization timescales are
thought to be very steep functions of the orbital semi-major axis
\citep{terquem98}, one could then argue that HAT-P-2b's large value of
$e$ is due to either the fact that the planet's orbit is not yet
circularized (while $\tau$~Boo~b's instead is), or to the presence of a
second planet in the HAT-P-2 system, or to rather different
formation/migration scenarios altogether.

\hd{147506}, with visual magnitude 8.71, is the fourth
brightest among the known stars harboring transiting planets. 
Therefore it has special interest because of the possibilities for
followup with large space or ground-based telescopes.

\acknowledgments

Operation of the HATNet project is funded in part by NASA grant
NNG04GN74G.
Work by G\'AB was supported by NASA through Hubble
Fellowship Grant HST-HF-01170.01-A.
GK wishes to thank support from Hungarian Scientific Research
Foundation (OTKA) grant K-60750.
We acknowledge partial support from the Kepler Mission under NASA
Cooperative Agreement NCC2-1390 (DWL, PI).
GT acknowledges partial support from NASA Origins grant NNG04LG89G.
TM thanks the Israel Science Foundation for a support through grant
no.~03/233.
The Keck Observatory was made possible by the generous financial
support of the W.~M.~Keck Foundation. DAF is a Cottrell Science
Scholar of Research Corporation, and acknowledges support from NASA grant
NNG05G164G.
We would like to thank Joel Hartman (CfA), Gil Esquerdo (CfA), Ron
Dantowitz and Marek Kozubal (Clay Center) for their efforts to observe
HAT-P-2b in transit, and Howard Isaacson (SFSU) for obtaining spectra
at Lick Observatory. We wish to thank Amit Moran for his help in the
observations with the Wise HAT telescope.
We owe special thanks to the directors and staff of FLWO, SMA and Wise
Observatory for supporting the operation of HATNet and WHAT.
We would also like to thank the anonymous referee for the useful
suggestions that improved this paper.




\end{document}